# Erosion yields of carbon under various plasma conditions in Pilot-PSI

K. Bystrov[a*], J. Westerhout[a], M. Matveeva[b], A. Litnovsky[b], L. Marot[c], E. Zoethout[a] and G. De Temmerman[a].

[a]*FOM Institute for Plasma Physics Rijnhuizen, Association EUROATOM-FOM, Trilateral Euregio Cluster, P.O. Box 1207, 3430 BE Nieuwegein, The Netherlands.*
[b]*Institut für Energieforschung - Plasmaphysik, Forschungszentrum Jülich GmbH, EURATOM Association, Trilateral Euregio Cluster, 52425 Jülich, Germany.*
[c]*University of Basel, Department of Physics, CH-4056 Basel, Switzerland.*

**Abstract.** Fine-grain graphite targets have been exposed to ITER divertor relevant plasmas in Pilot-PSI to address material migration issues in fusion devices. Optical emission spectroscopy and mass loss measurements have been employed to quantify gross chemical erosion and net erosion yields, respectively. Effects of the ion impact energy and target geometry on carbon erosion yields have been studied. It is concluded that temporal evolution of gross chemical erosion is strongly connected with changes in morphology of plasma exposed surfaces. The net carbon erosion yield is increased when the targets are partly covered by insulating boron-nitride rings.

## 1. Introduction

Successful demonstration of tokamak operation with burning plasma strongly depends on the lifetime of first wall and divertor materials. The lifetime of plasma-facing components is influenced by material erosion, subsequent transport of the released impurities through the plasma and their deposition. If the latter process occurs on plasma-exposed surfaces, deposited material will likely be re-eroded. On the contrary, material transport to remote areas leads to formation of co-deposited layers. These processes constitute material migration in fusion reactors [1]. ITER divertor materials will be subjected to high density, low temperature hydrogen plasmas ($n_e$ ~ $10^{21}$ m$^{-3}$, 1 eV $\leq T_e \leq$ 10 eV). Studies of processes related to material migration under such conditions are important from the point of view of determining the lifetime of plasma-facing components and tritium co-deposition in ITER divertor.

The present paper describes results from exposures of graphite targets under various plasma conditions in the Pilot-PSI linear device. Pilot-PSI is capable of producing plasmas with ITER divertor-relevant parameters [2]. Investigation of eroded carbon flux dependence on plasma electron temperature, surface temperature and ion flux are reported in [3]. Here, the focus is on studies of carbon erosion rates and re-deposition patterns as functions of ion impact energy, as well as target geometry and composition.

## 2. Experimental

2.1. Pilot-PSI and diagnostics

The Pilot-PSI device is described in detail in [2, 4]. The plasma source is a cascaded arc [5], which exhausts into the vessel along the magnetic field axis. The working gas in these experiments was hydrogen. Desired plasma conditions have been achieved by varying operating current, gas flow through the source and magnetic field.

Graphite targets of various diameters were clamped on a water cooled copper heat sink at a distance of 0.56 m from the source. The targets were at a floating potential or biased negatively with respect to the plasma potential. The surface temperature $T_{surf}$ in the centre of each target was recorded using a multi-wavelength pyrometer.

Radial profiles of plasma electron density $n_e$ and plasma electron temperature $T_e$ were measured with Thomson scattering (TS) system. A detailed description of the use of this diagnostic on Pilot-PSI can be found in [6].

Gross chemical erosion yield is determined *in-situ* by means of optical emission spectroscopy of hydrocarbon species [7, 8]. Light collected in front of the target is imaged on the entrance slit of the spectrometer optimized for spectroscopy of the molecular CH A – X band (band head around 431.42 nm).

2.2. Measurements of gross and net erosion yields.

In the present experiments both net and gross carbon erosion yields have been measured. The net erosion yield has been obtained by means of mass loss measurements, the target being weighed before and after the plasma exposure. The net erosion yield is thus:

$$Y_{net} = \frac{N_A \cdot \Delta m / M_C}{\Phi \cdot t}, \quad \text{[atoms/ion]} \tag{1}$$

where $N_A$ – Avogadro's number, $\Delta m$ – mass loss, $M_C$ – molecular mass of carbon, $\Phi$ – ion flux to the target, $t$ – exposure time. The incident ion flux is obtained by integrating the ion flux density over the target surface, assuming that ions in the sheath are accelerated up to the sound velocity and that plasma density in sheath drops by a factor of 2 compared to the pre-sheath. Alternatively, in cases of negatively biased targets, the ion saturation current was also used to determine the ion flux. Those two methods of calculating the ion flux yield same results within an accuracy of ~30%.

The CH signal intensity was determined by integrating the CH A – X band from 430.0 to 431.5 nm. The background was determined by measuring the intensity from 421.92 to 422.02 nm and 423.37 to 423.47 nm. In these regions the CH band emission is negligible and no Balmer lines are present. The optical emission spectrometer used for the gross erosion measurements was not calibrated absolutely. However, if the result shown in [8] demonstrating that the effective inverse photon efficiency for the CH A-X band depends weakly on $T_e$ is applied, the intensity of the CH signal can be interpreted as a direct measure of gross erosion:

$$Y_{gross}^{chem} = \frac{I_{CH}}{\Phi \cdot t} \quad \text{[a.u.]}, \qquad (2)$$

where $I_{CH}$ – intensity of the CH signal.

## 3. Results

A set of identical polished graphite targets (Ø 30 × 4 mm) were exposed to ~10 mm wide (width defined as FWHM of electron density profiles) plasma beam to monitor temporal evolution of the chemical erosion as a function of incident energy of hydrogen ions. In case of floating targets the impact energy is proportional to plasma electron temperature $E_i \sim 5T_e$ (assuming $T_i = T_e$ [9]). For biased targets the impact energy is a difference between plasma potential and the bias voltage. The plasma potential is estimated to be around -10 V. This value was obtained by subtracting potential drop in the sheath (~3$T_e$) from target's floating potential, which is measured to be close to -13 V. It was assumed that the plasma potential doesn't change when the bias is applied. Each target has been exposed to a series of plasma pulses. Plasma parameters, surface temperature and CH band emission were measured for each pulse. The flux variations were not exceeding 15% for each series of pulses. As shown in Fig. 1 (a), the temporal behavior of the chemical erosion of targets exposed to $T_e$ = 0.5 eV plasma differs strongly from the two other cases ($T_e$ = 1.3 eV and biasing). When $T_e$ = 0.5 eV, the CH A – X band emission intensity is independent of exposure time. For plasmas with higher $T_e$, emission intensity exhibits a non-linear temporal decrease. Simultaneously, the target surface temperature is found to increase with time (Fig. 1 (b)). It is important to mention that bulk temperature $T_{bulk}$, as measured by a thermocouple inserted into a target, doesn't follow the $T_{surf}$ behavior. On the contrary, $T_{bulk}$ doesn't change in time. It is well known that at surface temperatures well exceeding 1000 °C the formation rate of hydrocarbons is inferior to release of hydrogen from the surface [10]. A monotonic decrease of chemical erosion yield for increasingly high surface temperatures is shown in [11]. Thus, the decrease of the chemical erosion yield observed here might be attributed to the surface temperature increase.

Temperature increase is, in turn, related to the appearance of cauliflower-like deposits on the plasma exposed surfaces of the targets. Scanning electron microscopy (SEM) images reveal vast agglomerations of these dust particles, with their population density seemingly being maximal in the region of the peak particle flux ("top" image in Fig. 2). Surface profilometry confirms presence of deposits in plasma-wetted regions of graphite targets [3]. Due to their structure, these deposits have a bad thermal contact with the bulk material, leading to the mentioned increase of surface temperature. A somewhat similar effect has been observed in Tore Supra [12]. Fig. 2 also features a typical SEM picture of a target exposed to $T_e = 0.5$ eV plasma (images a,b). No signs of dust formation are present in this case, clearly indicating a link between changes of surface morphology and plasma conditions. Dust formation on the surface of biased targets is a topic of ongoing studies.

The observed dust formation does not prevent net erosion of carbon as shown in Fig. 3, since the measured net erosion yield increases with the ion energy. Obtained dependence of the net erosion yield on ion impact energy is consistent with results of calculations in [13], taking into account ion energies $E_{ion} < 100$ eV and surface temperatures exceeding 1000 °C. Interestingly enough, net erosion yields of diamond coatings exposed in Pilot-PSI in continuation of work reported in [14] are similar to those of carbon. However, it must be pointed out that surface temperatures of carbon and diamond samples were very different. Surface temperatures of carbon were exceeding 1000 °C, while those of diamond targets stayed below the detection level of the pyrometer (around 600 ºC for the exposure time used). Still, erosion yields of both materials follow the same energy dependence.

To investigate influence of the target geometry and composition on erosion yields a dedicated set of small (Ø 5, 10 and 15 mm) graphite targets clamped into protective rings has been exposed. Dimensions of the targets have been selected to be smaller, equal to, or larger than the plasma beam diameter, to get some insight into where re-deposition occurs on the target. The rings have been manufactured either out of tungsten, or out of boron nitride. Results of the exposures are presented in Fig. 4. Surface temperatures of all samples were in the range of 1200-1300 ºC. Firstly, lower net erosion yield for larger targets suggests that a fraction of eroded material is re-deposited on the target surface. Secondly, it is important to notice that the targets inserted into boron-nitride insulating rings show higher net erosion yields compared to those inserted into tungsten rings. Boron nitride rings were themselves eroded as confirmed by mass loss measurements. Thus, nitrogen and boron are released into the plasma during exposures. Consequently, the plasma impinging on the target contains those two impurities in addition to hydrogen. Since nitrogen can chemically sputter carbon [15], it is not unreasonable to speculate on its influence on the results described here. Further experiments are foreseen to quantify the influence of these impurities.

**4. Conclusions**

During exposures of carbon targets to ITER divertor relevant plasmas in Pilot-PSI a strong influence of ion impact energy on temporal evolution of gross chemical erosion has been observed. Decrease of the CH A – X band signal is accompanied by appearance of cauliflower-like particles on plasma exposed areas of the targets. Despite the dust formation, the net erosion yield of carbon is increasing with ion impact energy in the range used in these experiments ($E_{ion} \leq 75$ eV). Experiments with targets of various sizes show that presence of nitrogen and boron in Pilot-PSI plasma increases the net erosion yield.

**Acknowledgement**

The authors would like to thank R.S. Al, M.J. van de Pol, H.J. van der Meiden, M. Zlobinski, J.J. Zielinski, A.E. Shumack, W.A.J. Vijvers, G.M. Wright and G.J. van Rooij for their support in preparation and execution of the experiments. This work was supported by the European Communities under the contract of Association between EURATOM/FOM and carried out within the framework of the European Fusion Programme with financial support from NWO and the NWO Grant No. RFBR 047.018.002. The views and opinions expressed herein do not necessarily reflect those of the European Commission.

Figure captions

Fig. 1. (a) Temporal evolution of gross chemical erosion for targets exposed to plasmas with different parameters. (b) Corresponding temporal evolution of surface temperatures. For floating targets values of plasma electron temperature are given.

Fig. 2. Characteristic SEM images of a surface where no dust has been deposited (a,b) and a surface covered with cauliflower-like dust structures (c,d).

Fig. 3. Net erosion yield of carbon and boron-doped diamond as a function of ion incident energy. Surface temperature of the graphite samples was $T_{surf} > 1000$ ºC, while that of the boron-doped diamond was $T_{surf} < 600$ ºC.

Fig. 4. Influence of the target size and material of the surrounding protective ring on net erosion yield.

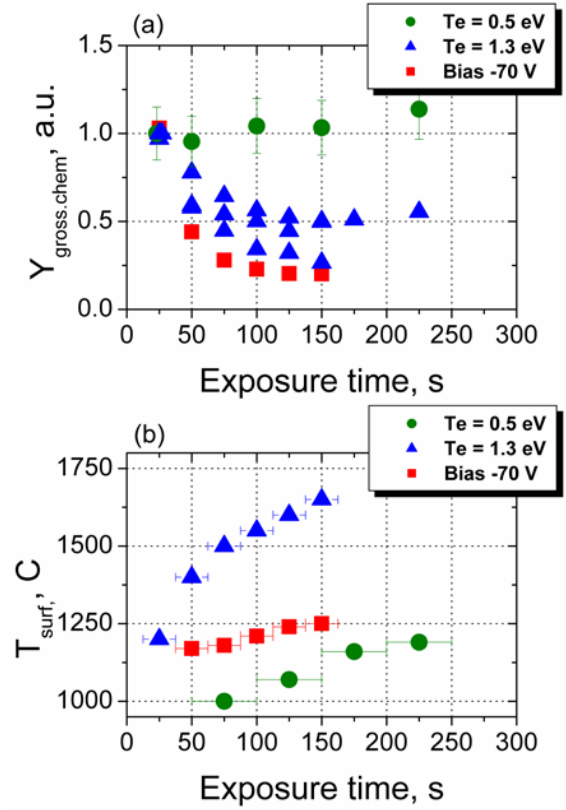

Fig 1. (a) Temporal evolution of gross chemical erosion for targets exposed to plasmas with different parameters. All data series are normalized to unity. (b) Corresponding temporal evolution of surface temperatures. For floating targets values of plasma electron temperature are given.

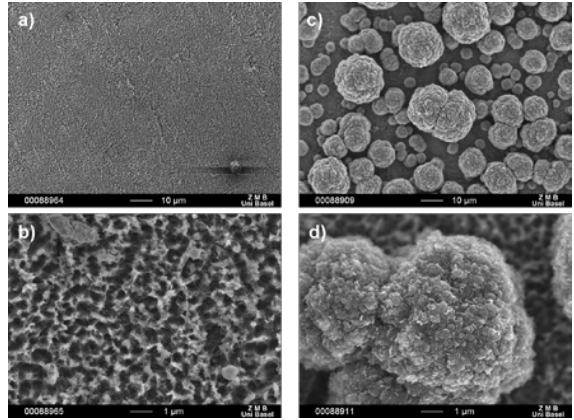

Fig. 2. Characteristic SEM images of a surface where no dust has been deposited (a,b) and a surface covered with cauliflower-like dust structures (c,d).

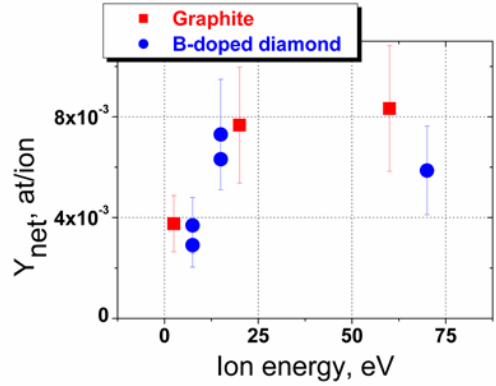

Fig. 3. Net erosion yield of carbon and boron-doped diamond as a function of ion incident energy. Surface temperature of the graphite samples was $T_{surf} > 1000$ °C, while that of the boron-doped diamond was $T_{surf} < 600$ °C.

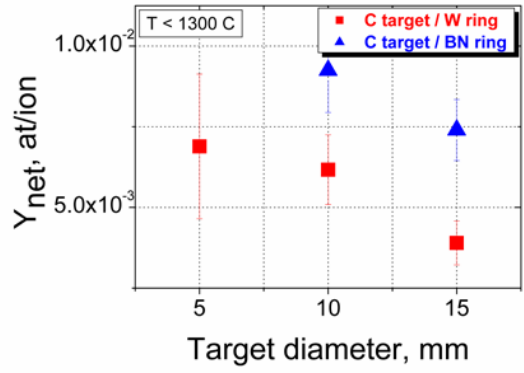

Fig. 4. Influence of target size and material of the surrounding protective ring on net erosion yield.